\documentclass[nofootinbib,prd,aps,amsmath,superscriptaddress,tightenlines]{revtex4}

\newcommand{\nn}{\nonumber}
\newcommand{\gb}{\bar{g}}
\newcommand{\nb}{\overline{\nabla}}
\newcommand{\rb}{\overline{R}}

\begin{document}


\preprint{ \vbox{\hbox{UCSD/PTH 02-13}\hbox{UTPT 02-06} \hbox{hep-ph/0206212}  }}

\title{Gauge Fields and Scalars in Rolling Tachyon Backgrounds}

\author{Thomas Mehen}
\affiliation{Department of Physics, Duke University, Durham NC 27708 \\
	Jefferson Laboratory, 12000 Jefferson Ave. Newport News VA 23606\footnote{Electronic address: mehen@phy.duke.edu}}

\author{Brian Wecht}
\affiliation{Department of Physics, UC San Diego, 
	La Jolla, CA 92093\footnote{Electronic address: bwecht@physics.ucsd.edu}}

\date{\today\\
\vspace{.3cm} 
}

\begin{abstract}

We investigate the dynamics of gauge and scalar fields on unstable D-branes with rolling tachyons. Assuming an FRW metric
on the brane, we find a solution of the tachyon equation of motion which is valid for arbitrary tachyon potentials and
scale factors. The equations of motion for a U(1) gauge field and a scalar field in this background are derived.  These
fields see an effective metric which differs from the original FRW metric. The field equations receive large corrections
due to the curvature of the effective metric as well as the time variation of the gauge coupling. The equations of state
for these fields resemble those of nonrelativistic matter rather than those of massless particles. 

\end{abstract}

\maketitle

\section{INTRODUCTION}

Tachyons arise generically in string theory whenever there is no space-time supersymmetry. In the bosonic theory all
branes have tachyons, and in superstring theory they appear as modes on non-BPS branes or brane-antibrane pairs. The
universe is non-supersymmetric and non-static; this has prompted speculation that it may be described by a non-BPS
configuration that is evolving in time, possibly toward a supersymmetric vacuum. Recently, Sen \cite{Sen2} has proposed
studying the cosmological implications of tachyons whose action is given by
\begin{eqnarray}\label{OriginalAction}
{\cal S} = - \int\!\! d^{p+1}x \, V(T)\sqrt{-\det\left(g_{\mu \nu} + \partial_\mu T \partial_\nu T\right)} \, ,
\end{eqnarray}
where $T$ is the tachyon field, $g_{\mu \nu}$ is the metric, and  the tachyon potential $V(T)$ has an unstable
maximum at $T=0$ and a minimum at $T = \infty$.  This form of the action for tachyons on bosonic or non-BPS branes
has been verified by explicit calculations \cite{Garousi,Bergshoeff,Kluson,AFTT}. 

String theory tachyons have several unusual features which distinguish them from ordinary  scalar fields. The tachyon field
equations do not have plane wave solutions \cite{Sen3}. (When Eq.~(\ref{OriginalAction}) is extended to include gauge fields
there are no plane wave solutions for these fields \cite{Ishida} either.) Furthermore, a rolling tachyon cannot oscillate about the
minimum of its potential  like a scalar whose potential minimum lies at some finite value of the field. The endpoint of the
tachyon's evolution corresponds to the decay of the unstable brane, or the annihilation of the brane-antibrane pair. The open
string degrees of freedom  should disappear as $T$ rolls to the minimum of $V(T)$. The absence of oscillations around the
minimum of the potential is consistent with this expectation. The energy-momentum tensor of a homogeneous tachyon field is
that of a perfect fluid with energy density

\begin{eqnarray}\label{energydensity}
\rho = \frac{V(T)}{\sqrt{1-\dot{T}^2}}
\end{eqnarray}
and pressure
\begin{eqnarray}\label{pressure}
p = -V(T)\sqrt{1-\dot{T}^2}\, ,
\end{eqnarray}
implying that the equation of state for the tachyon fluid is
\begin{eqnarray}\label{EOS}
p = -(1-\dot{T}^2)\rho \,.
\end{eqnarray}
For $\dot{T}=0$, the equation of state is that of vacuum energy, and for $\dot{T}=1$, it is the equation of state of
nonrelativistic matter. As the tachyon rolls, the equation of state will interpolate  between these two extremes.
For this reason the tachyon has been investigated as a candidate for the inflaton, dark matter, and
quintessence \cite{Fairbairn,Mukohyama,Feinstein,Choudhury,Linde,Sami,Sami2,Minahan2,Cornalba,Hashimoto,Kim,Frolov,Shiu,
Padmanabhan2,Sen1,Gibbons1,Padmanabhan1,Li,Sugimoto,Benaoum,Hwang}. 

Problems with identifying the tachyon as the inflaton are discussed in Ref.~\cite{Linde}, where it is argued that there is
no obvious mechanism for reheating the universe after tachyon driven inflation because of the absence of oscillations
about the minimum of the potential. Therefore, one expects the tachyon to dominate the energy density of the universe
after inflation, and since in this phase the tachyon energy density falls off as $a^{-3}$, tachyon energy density  would
dominate over radiation. On the other hand, it is known that the early universe went through a radiation dominated phase.
This argument does not exclude the possibility of tachyon driven inflation followed by an inflationary era driven by
another field \cite{Linde}. The tachyon could contribute to dark matter, though it requires fine-tuning of initial
conditions to ensure that the present tachyon energy density is not so large as to exceed current bounds on dark matter or
too small to make a significant contribution \cite{Shiu}.

It is worthwhile to point out that the above conclusions are based on classical solutions of Eq.~(\ref{OriginalAction}).
These solutions correctly mimic the late time behavior of solutions of open string field theory found in Ref. \cite{Sen1}.
These solutions were obtained by deforming the CFT of an unstable D-brane by an exactly marginal operator and therefore are
exact classical solutions of weakly coupled open string theory. However, the actual behavior of an unstable D-brane in 
string theory could be drastically different due to quantum effects. In particular, Ref.~\cite{OS} has shown that the
coupling of unstable D-branes to massive closed string states grows expontentially as the brane decays. The non-decoupling of
massive string states would invalidate  Eq.~(\ref{OriginalAction}) as a low energy effective field theory for
late time behavior of an unstable D-brane in string theory.  Most of the recent work on cosmology of tachyonic matter has
ignored possible non-decoupling effects of massive closed string states and instead taken the action of 
Eq.~(\ref{OriginalAction}) as a starting point of the analysis.

In scenarios in which an open string theory tachyon contributes to dark matter or plays the role of an inflaton,  
it is natural
to ask what happens to the gauge fields and other massless fields on the decaying brane as the tachyon rolls to its 
minimum. For these
scenarios, it is important to know whether the massless fields on the brane could be identified with those of 
the Standard Model. The
late time behavior of gauge and scalar fields living on an unstable brane is also interesting in its own right. 
In this paper, we
find a solution for a homogeneous tachyon field in Friedman-Robertson-Walker (FRW) cosmological backgrounds for arbitrary $V(T)$. This
solution is valid for $\dot{T} \approx 1$, i.e.\ when the tachyon equation of state is like that of nonrelativistic matter. We then
add scalar and gauge fields to the action in Eq.~(\ref{OriginalAction}) in the following manner,
\begin{eqnarray}\label{ActionwithF}
{\cal S} &=& -\int\!\! d^{p+1}x\, V(T)\sqrt{-\det\left[g_{\mu \nu} + 
\partial_\mu T \partial_\nu T + 2\pi l_s^2( F_{\mu \nu} + \partial_{\mu} \phi \partial_{\nu} \phi ) \right ] } \, ,
\end{eqnarray}
where $l_s$ is the string length (for the remainder of this paper, we will set $2\pi l_s^2 = 1$). This form of the action
for a non-BPS brane of string theory has been checked by explicit calculations of S-matrix elements \cite{Garousi} and is
consistent with T-duality \cite{Bergshoeff}. From the action in Eq.~(\ref{ActionwithF}), we derive equations of  motion for
the gauge and scalar fields which are valid for small fluctuations and do not include the backreaction of the tachyon and
the metric.  The scalar and gauge fields see an effective metric different from the metric seen by gravity.  Furthermore,
the equations of state of scalar and gauge fields differ from the ordinary equations of state. In both cases the pressure
is suppressed by factors of $1-\dot{T}^2$, so their contribution to the energy-momentum tensor is like that of cold dark
matter rather than  ordinary massless fields.  Finally, the gauge coupling grows rapidly with time in these models.
Therefore, it is extremely difficult to identify gauge fields living on a non-BPS brane 
with Standard Model fields.

The results of this paper should be of relevance for string theory studies of tachyon condensation. Most studies focus on
D-branes decaying in flat space-time. Our analysis considers a brane decaying in a time dependent gravitational
background.  Many authors \cite{Yi,Bergman,Kleban,Sen4,Gibbons2} have argued that the gauge theory living on a D-brane becomes
strongly coupled as the brane decays and that the gauge fields are confined at the endpoint of tachyon condensation. In
our solutions the gauge coupling is indeed seen to grow with time. Ref.~\cite{Sen2} found time dependent solutions for
tachyons in string field theory and showed that the decaying D-brane  evolves into a pressureless gas at late time. This
analysis considered only the behavior of the tachyon; our analysis shows that contributions to the pressure from gauge
fields and scalars also vanish at late times. It would be interesting to confirm this in a full string field theory
calculation. In addition, the equations of motion for the gauge fields and scalars show that these fields see an effective
metric whose curvature is growing rapidly with time. This introduces large curvature corrections to the equations of
motion for these fields. It is important to emphasize that these corrections are absent unless both the tachyon and the
metric are evolving with time. These large curvature corrections are a novel feature of tachyon condensation which have
not been discussed previously in the literature. 

\section{SOLUTION OF THE TACHYON EQUATION OF MOTION}

In an FRW universe with metric
\begin{eqnarray}\label{FRW}
ds^2 = -dt^2 + a^2(t)\left(\frac{dr^2}{1- kr^2} +r^2 d\Omega^2\right) \,,
\end{eqnarray}
the equation of motion for a spatially homogeneous tachyon is
\begin{eqnarray}\label{Teqn}
\frac{\ddot{T}}{1-\dot{T}^2} + 3\frac{\dot{a}}{a}\dot{T} + \frac{V^\prime(T)}{V(T)} = 0 \, ,
\end{eqnarray}
where $^\prime$ denotes differentiation with respect to $T$.
The scale factor obeys the equation
\begin{eqnarray}\label{Heqn}
\left ( \frac{\dot{a}}{a} \right )^2 = \frac{8\pi G}{3}\rho  -\frac{k}{a^2} \,,
\end{eqnarray}
where $\rho$ includes all contributions to the energy density. We will solve Eq.~(\ref{Teqn}) directly
in terms of $a(t)$, so the solution will be valid for arbitrary matter content and curvature. Solutions 
with finite energy density must have $\dot{T}\rightarrow 1$ as $V(T) \rightarrow 0$. Therefore, we look for a
solution of the form
\begin{eqnarray}
T(t) = t + f(t)\,,
\end{eqnarray}
where $f(t)$ is assumed to be a small correction. Solutions of this form for the specific case of  $V(T) \propto
e^{-T}$ are known for flat space \cite{Sen3} and for the FRW metric \cite{Frolov}. We generalize these solutions to any
$V(T)$. Using the approximations $f(t) \ll t$ and  $\dot{f}(t) \ll 1$, the equation for $f$ is 
\begin{eqnarray}
-\frac{\ddot{f}}{2 \dot{f}}+3 \frac{\dot{a}}{a} + \frac{\dot{V}}{V} = 0
\end{eqnarray}
This equation is easily solved since every term is a total time derivative, yielding
\begin{eqnarray}\label{soln}
\dot{T}(t) &=& 1- \frac{1}{2\rho_T^2} \, a^6(t)V^2(t)  \,,\\
T(t) &=& t-t_i + T(t_i) - \frac{1}{2\rho_T^2} \, \int_{t_i}^t dt^\prime a^6(t^\prime) V^2(t^\prime) \,. \nonumber
\end{eqnarray}
The solution in Eq.~(\ref{soln}) has two constants of integration, one of which is expressed in terms of $\rho_T$, the
tachyon energy density when $a(t) = 1$, the other in terms of the value of the tachyon field at some initial  time $t_i$. In
most cosmological applications, it is\ assumed that the tachyon starts  rolling at some early time $t_{\rm R}$ with
$T=0$ and $\dot{T}=0$. The solution in Eq.~(\ref{soln}) is not valid all the way back to $t_{\rm R}$. Therefore, we 
must take $t_i > t_{\rm R}$, though for a physically sensible solution one expects that the
condition $\dot{T}\approx 1$ should be satisfied shortly after the tachyon begins its descent. The energy
density and pressure of the tachyon fluid as a function of time are
\begin{eqnarray}
\rho(t) &=& \frac{\rho_T}{a^3(t)} \,,\\
p(t) &=&  -\frac{V^2 a^3(t)}{\rho_T} = - \frac{V^2(t) a^6(t)}{\rho_T^2}\rho (t) \,.\nonumber
\end{eqnarray}
 
The approximations $f(t) \ll t$ and  $\dot{f}(t) \ll 1$ are valid at late times if $a^3(t) V(t) \rightarrow 0$   as
$t \rightarrow \infty$. Though the functional form of $V(T)$ is not known, it is expected to decay exponentially as a
function of $T$.  Two functional forms which satisfy the requirements that the potential have an unstable maximum at
$T=0$ and a minimum at $T=\infty$ are
\begin{eqnarray}\label{sft}
V(T) &=& T_p \left(1+\frac{T}{T_0}\right) e^{-T/T_0} \qquad {\rm bosonic} \,, \\
 &=&  T_p \, e^{-T^2/T_0^2} \qquad \qquad \qquad {\rm superstring} \,.\nn
\end{eqnarray}
$T_p$ is the p-brane tension, which has mass dimension 4 for $p=3$. These potentials are similar to tachyon potentials
found in bosonic \cite{Kutasov1,Gerasimov,Tseytlin} and supersymmetric \cite{Tseytlin,Kutasov2} string  theory
calculations. The actions appearing in these calculations are of a slightly different form than
Eq.~(\ref{OriginalAction});  in particular, the tachyon kinetic terms do not appear inside a square root. If one expands
the action of Eq.~(\ref{OriginalAction}) in powers of $\partial T$, the result is very similar to the actions in
\cite{Kutasov1,Gerasimov,Tseytlin,Kutasov2}. Recently, Sen has proposed $V(T) = T_p\exp(-|T|/T_0)$ \cite{Sen3} for
$V(T)$. For these potentials, the pressure vanishes exponentially at late times.

\section{DYNAMICS OF GAUGE FIELDS AND SCALARS}

In this section, we study the dynamics of gauge fields and scalars in a rolling tachyon background using the action of
Eq.~(\ref{ActionwithF}). The (abelian) field strength is included in a manner consistent with T-duality, and the
inclusion of scalar fields (which describe the transverse motion of the brane) arises naturally from considering the
transverse components of the induced metric. There is one such scalar field for each direction normal to the brane. We
have included only one scalar field here for simplicity; our analysis is easily extended to any number of these scalars.
The form of the action for a non-BPS D-brane has been a subject of intense study. It has been checked by explicit
calculations of $S$ matrix elements and is consistent with T duality \cite{Garousi,Bergshoeff,Kluson,AFTT}.

Although we are primarily interested in small scalar and gauge field fluctuations in 
the time-dependent background, it is not difficult to find the exact equations of motion. 
The action of Eq.~(\ref{ActionwithF}) can be written as
\begin{eqnarray}
{\cal S} = - \int \! \! d^{p+1}x V(T) \sqrt{- \det A}
\end{eqnarray}
where $A_{\mu \nu} = g_{\mu \nu} + \partial_{\mu} T \partial_{\nu} T + 
\partial_{\mu}\phi \partial_{\nu}\phi + F_{\mu \nu}$, one finds
\begin{eqnarray}\label{exactgf}
\partial_{\mu} \left [\frac{1}{2}V(T)\sqrt{-\det A}\left [ (A^{-1})^{\mu \nu} - (A^{-1})^{\nu \mu} \right ] \right ] = 0
\end{eqnarray}
for the equation of motion of the gauge field and
\begin{eqnarray}
\partial_{\mu} \left [\frac{1}{2}V(T)\sqrt{-\det A}\left [ (A^{-1})^{\mu \nu} + (A^{-1})^{\nu \mu} \right ]
\partial_{\nu} \phi \right ] = 0
\end{eqnarray}
for the equation of motion of the scalar field. Similarly, the exact equation of motion for the tachyon is
\begin{eqnarray}
V^{\prime} (T) \sqrt{-\det A} = \partial_{\mu} \left [\frac{V(T)}{2}\sqrt{-\det A}
\left [(A^{-1})^{\nu \mu} + (A^{-1})^{\mu \nu}\right ] \partial_{\nu} T  \right ].
\end{eqnarray}
For completeness, we also include here the equation for conservation of the stress tensor,
\begin{eqnarray}
\nabla_{\mu}\left [-\frac{1}{2}V(T)\sqrt{\frac{A}{g}}\left [ (A^{-1})^{\mu \nu} + (A^{-1})^{\nu \mu} \right ] \right ] = 0.
\end{eqnarray}

We now specialize to the case of a 3+1 dimensional brane and study the dynamics of small scalar and gauge field
fluctuations. For this purpose, we expand the action of Eq.~(\ref{ActionwithF}) to quadratic order in the fields $F_{\mu \nu}$
and $\phi$.  For related investigations of Born-Infeld dynamics, see \cite{Gibbons3,Gibbons4,Gibbons5}. First, we set
$\partial_{\mu} \phi = 0$ and find the form of the action for the U(1) gauge field. We can simplify the analysis by
introducing the effective metric
\begin{eqnarray}
\gb_{\mu \nu} = g_{\mu \nu} + \partial_{\mu} T \partial_{\nu} T,
\end{eqnarray}
 and using the identity (valid in four dimensions)
\begin{eqnarray}
- {\rm det}(\gb_{\mu \nu} + F_{\mu \nu}) =
- \gb\left(1+ \frac{1}{2}\gb^{\alpha \beta}\gb^{\gamma \delta}F_{\alpha \gamma}F_{\beta \delta} \right) 
- {\rm det}(F_{\mu \nu})
\end{eqnarray}
where $\gb ={\rm det}(\gb_{\mu \nu})$ and $\gb^{\mu \nu}$
is defined by $\gb^{\alpha \mu}\gb_{\mu \beta} = \delta^{\alpha}_{\,\,\, \beta}$. 
$\gb_{\mu \nu}$ is the metric seen by gauge fields, i.e. the open string metric. The fact
that this is different from the closed string metric is widely appreciated; see for example \cite{Gibbons4}.
In this paper, we explore the effect of a rolling tachyon on the open string metric.
As is well known, there is a critical value $E_{cr}$ for an electric field on a D-brane. For
$E \equiv |F_{0i}|> E_{cr}$ the Dirac-Born-Infeld action no longer makes sense; this corresponds to the D-brane
being destroyed by the strong electric field. When magnetic fields are vanishing, $F_{ij}=0$, 
$E_{\rm cr} = a \sqrt{1-\dot{T}^2}$.

We now derive the equations of motion for small field fluctuations satisfying $E \ll E_{cr}$.
Therefore, we can expand the Lagrangian to lowest order in $F^2$. This yields
\begin{eqnarray}\label{lag}
{\cal L} = -V(T)\sqrt{-\gb} -
 V(T)\frac{1}{4}\sqrt{-\gb}\, \gb^{\alpha \beta}\gb^{\gamma \delta}F_{\alpha \gamma}F_{\beta \delta}
\end{eqnarray}
The form of the Lagrangian for a scalar field is similar; expanding to 
lowest order in $(\partial \phi)^2$ gives
\begin{eqnarray}
{\cal L}_{\phi} = -V(T)\frac{1}{2}\sqrt{-\gb}\gb^{\mu \nu}\partial_{\mu} \phi \partial_{\nu} \phi.
\end{eqnarray}
Since we have assumed both $\partial \phi$ and $F_{\mu \nu}$ are small we have ignored all 
coupling terms between the two fields, which would be order $(\partial \phi)^2 F^2$ or higher.

Deriving the equations of motion is straightforward. For the scalar field, one finds 
\begin{eqnarray}\label{phieq}
\left [ \nb_{\mu} + (\partial_{\mu}\ln V) \right ] (\gb^{\mu \nu}\partial_{\nu} \phi) = 0
\end{eqnarray}
where $\nb_{\mu}$ is the covariant derivative with respect to the metric $\gb_{\mu \nu}$. The Christoffel
symbols and curvature tensors for the metric $\gb_{\mu \nu}$ for a spatially homogeneous
$T$ in an FRW background with $k=0$ are calculated in the Appendix.
We can simplify this equation 
further by defining $g^{\prime}_{\mu \nu} = V(T) \gb_{\mu \nu}$, which yields
\begin{eqnarray}
\nabla^{\prime}_{\mu} \nabla^{\prime \mu} \phi = 0
\end{eqnarray}
where the covariant derivative is defined with respect to the metric $g_{\mu \nu}^{\prime}$. This result could have been
derived by first rescaling the metric in the Lagrangian, which becomes ${\cal L}_{\phi} =
-\frac{1}{2}\sqrt{-g^{\prime}}g^{\prime \mu \nu}\partial_{\mu} \phi \partial_{\nu} \phi$. 

If we consider the case of a flat universe (k=0) we can use the formulae in the Appendix to rewrite Eq.~(\ref{phieq}) as
\begin{eqnarray}\label{phisimple}
\gb^{\mu \nu} \partial_\mu \partial_\nu \phi - 3\frac{\dot{a}}{a}\dot{\phi} =0 \,.
\end{eqnarray}
The second term is suppressed by $M_{pl}$ since $\dot{a}/{a} \sim \sqrt{ G \rho} \sim \sqrt{\rho}/M_{pl}$. Dropping this
term, we see that $\phi$ has solutions which propagate along null geodesics of the metric $\gb_{\mu \nu}$ instead of
$g_{\mu \nu}$. Note that null geodesics of $g^{\prime}_{\mu \nu}$ are the same as null geodesics of $\gb_{\mu \nu}$ since
the metrics differ only by a conformal factor. Thus, the scalar fields see an effective metric $g^{\prime}_{\mu \nu} =
V(T) ( g_{\mu \nu} + \partial_{\mu} T \partial_{\nu} T)$ which is not the metric seen by gravity, $g_{\mu \nu}$.

The source-free Maxwell equation derived from the Lagrangian 
in Eq.~(\ref{lag}) is 
\begin{eqnarray}\label{Maxwell}
\left [ \nb_{\alpha} + (\partial_{\alpha}\ln V)\right ]F^{\alpha \mu} = 0 \,,
\end{eqnarray}
where $F^{\alpha \mu} \equiv \gb^{\alpha \beta} \gb^{\mu \nu} F_{\beta \nu}$. We can write this in the form of a wave
equation for the gauge field by choosing the gauge $\nb_{\alpha} A^{\alpha} = 0$,
\begin{eqnarray}\label{vecpot}
\nb_{\alpha}\nb^{\alpha} A^{\mu} - \rb_{\beta}^{\,\,\,\, \mu}A^{\beta} + 
(\partial_{\alpha} {\rm ln}V)(\partial^{\alpha} A^{\mu} - \partial^{\mu}A^{\alpha}) = 0 \, .
\end{eqnarray} 
It is also possible to derive a wave equation for the field strengths by combining Eq.~(\ref{Maxwell}) with the Bianchi
identity:
\begin{eqnarray}\label{fs}
\nb_{\alpha}\nb^{\alpha}F^{\beta \gamma} + 2\rb^{\beta \,\,\,\,\,\,\, \gamma}_{\,\,\, \alpha \mu}F^{\mu \alpha}
+\rb_{\mu}^{\,\,\,\, \gamma}F^{\mu \beta} - \rb_{\mu}^{\,\,\,\, \beta}F^{\mu \gamma}
+\nb^{\beta}((\partial_{\alpha}\ln V)F^{\alpha \gamma}) - \nb^{\gamma}((\partial_{\alpha}\ln V)F^{\alpha \beta})= 0.
\end{eqnarray}
The bars over the covariant derivatives and curvature tensors indicate that they are defined  with respect to the metric 
$\gb_{\mu \nu}$. These equations describe an electromagnetic field propagating in a space-time with metric $\gb_{\mu \nu}$
and a time-varying coupling constant  $g^2 \sim 1/V(t)$. This can also be seen directly from the Lagrangian in
Eq.~(\ref{lag}). The corrections from the curvature and  time dependent coupling are large. For  $V(T) \sim e^{-T/T_{0}}$,
$|\partial_{0}\ln V| \sim 1/T_{0} \sim M_{pl}$. Here we assume all dimensional quantities in the Lagrangian have their
scale set by $M_{pl}$. Curvature corrections are suppressed by $M_{pl}$ when $\dot{T}\approx 0$ but are large
at later times because all nonzero components of the Ricci  tensor $\rb_{\mu}^{\,\,\,\, \nu}$ have a prefactor of
$1/(1-\dot{T}^2)$, which grows exponentially as the  tachyon rolls to its minimum. Unlike the case of scalars, we have not
attempted to rewrite the equations for electromagnetic waves in the form of a simple wave equation like
Eq.~(\ref{phisimple}).  Therefore, it is not  clear whether electromagnetic waves will propagate along null 
geodesics of the metric $\gb_{\mu \nu}$.

As a simple example, we can solve Eq.~(\ref{Maxwell}) for a spatially homogeneous electric field $E \equiv F_{0i}$.
Using the results in the Appendix it is straightforward to show that
\begin{eqnarray}\label{ConstantE}
\partial_{0}E = - \left [ \frac{\dot{T}\ddot{T}}{1-\dot{T}^2} + \frac{\dot{a}}{a} + \frac{V^\prime}{V}\dot{T} \right ] E.
\end{eqnarray}
The solution of Eq.~(\ref{ConstantE}) is given by
\begin{eqnarray}\label{et}
E = E_0\left [ \frac{\sqrt{1-\dot{T}^2}}{aV} \right ] = \frac{E_0}{\rho_T}a^2
\end{eqnarray}
where $E_0$ is a constant. One expects the scale factor to grow with time, so our solution will break down when $E \sim
E_{cr}$. Since $E_{cr} = a \sqrt{1-\dot{T}^2}$,  our solution will be valid until $a^2 V \sim E_{0}$. To
understand what happens in this regime, one would have to take into account the corrections  from terms that
are $O(F^4)$ and higher as well as corrections to the equations of
motion of the tachyon and metric  from the electric field.  It is
easy to see that the growth of the electric field in excess of $E_{cr}$ is an artifact of truncating the action
to $O(F^2)$. The exact energy density of a brane with a uniform electric field is given by 
\begin{eqnarray}
\rho = \frac{V(T)}{\sqrt{1 - \dot{T}^2 - (E^2/a^2)}} \,.
\end{eqnarray}
We can rewrite this equation in terms of the critical field $E_{cr}$ as 
\begin{eqnarray}\label{eth}
E_{cr}^2 - E^2 = \frac{V^2 a^2}{\rho^2} \, .
\end{eqnarray}
Since the right hand side is positive definite, $E \leq E_{cr}$. 

Next, we compute the $O(F^2)$ and $O(\phi^2)$ corrections to the energy-momentum tensor. It is of the form 
\begin{eqnarray}
T_{\mu \nu} = T^{T}_{\mu \nu} + T^{F}_{\mu \nu} + T^{\phi}_{\mu \nu}
\end{eqnarray}
where the contribution from tachyons is \cite{Frolov}
\begin{eqnarray}
T^{T}_{\mu \nu} = V(T)\left ( \frac{\partial_{\mu} T \partial_{\nu} T}{\sqrt{1 + \partial_{\beta} T \partial^{\beta}T}} 
- g_{\mu \nu}\sqrt{1 + \partial_{\beta} T \partial^{\beta}T}\right ),
\end{eqnarray}
the $O(F^2)$ term is 
\begin{eqnarray}\label{tf}
T^{F}_{\mu \nu} = \frac{1}{4} T^{T}_{\mu \nu} \,\gb^{\alpha \beta}\gb^{\gamma \delta}F_{\alpha \gamma}F_{\beta \delta} 
+ V(T) (\gb g)^{\alpha}_{\,\,\,\, \mu}( \gb g )^{\beta}_{\,\,\,\,  \nu} 
\gb^{\gamma \delta} F_{\alpha \gamma} F_{\beta \delta}\sqrt{1 + \partial_{\sigma} T \partial^{\sigma}T} \, ,
\end{eqnarray}
and the $O(\phi^2)$ term is
\begin{eqnarray}
T^{\phi}_{\mu \nu} =\frac{1}{2}T^{T}_{\mu \nu} \,\gb^{\alpha \beta}\partial_{\alpha}\phi \partial_{\beta} \phi 
+ V(T) (\gb g)^{\alpha}_{\,\,\,\, \mu}( \gb g )^{\beta}_{\,\,\,\,  \nu} \partial_{\alpha}\phi \partial_{\beta} \phi 
\sqrt{1 + \partial_{\sigma} T \partial^{\sigma}T} \,.
\end{eqnarray}
Here $(\gb g)^{\alpha}_{\,\,\,\, \mu} = \gb^{\alpha \beta} g_{\beta \mu}$ and 
$\partial_{\beta}T \partial^{\beta}T \equiv g^{\alpha \beta}\partial_{\alpha} T \partial_{\beta} T$.
The $O(F^2)$ piece contains contributions from both the tachyon and gauge fields and is not of the conventional
form for gauge fields. In particular $g^{\mu \nu}T^F_{\mu \nu} \neq 0$, so the contribution of $T^F_{\mu \nu}$ to the equation of
state differs from conventional radiation. It is possible
to put $T^F_{\mu \nu}$ in perfect fluid form by assuming the radiation is isotropic; we then find 
\begin{eqnarray}
p_F = \frac{1}{3} (1 - \dot{T}^2) \rho_F.
\end{eqnarray}
Again the pressure of the radiation is vanishing as $\dot{T}\rightarrow1$ so gauge fields on the brane
behave like nonrelativistic matter rather than conventional radiation. For a spatially homogeneous scalar field, the
equation of state is 
\begin{eqnarray}
p_{\phi} = (1-\dot{T}^2)\rho_{\phi},
\end{eqnarray}
which also has a pressure that vanishes as $\dot{T} \rightarrow 1$.

\section{Conclusions}

Gauge fields and scalars living on a D-brane with a rolling tachyon have very different physics from Standard Model fields
in conventional cosmological scenarios.  The massless fields on a decaying D-brane see different metrics than the one seen
by gravity. There are large corrections to the classical equations of motion from the curvature of this effective metric
as well as the time variation of the gauge coupling.   The equations of state for both the gauge field and the scalar are
also non-standard, with  a pressure that vanishes as the tachyon rolls to the minimum of its potential. These significant
departures from the physics of normal gauge and scalar fields make it impossible to identify Standard Model fields with
the fields living on a non-BPS brane whose action is of the form in Eq.~(\ref{ActionwithF}). 

This analysis reveals interesting features of the process of brane decay.  Our calculations confirm that the gauge
fields become strongly coupled as the tachyon rolls to its minimum. Furthermore, we showed that the contribution of gauge
fields and scalars to the pressure vanishes at late times. We also found  large corrections to the equations of motion for
massless fields on the brane because the curvature of the effective metric is growing rapidly with time. These corrections
are only seen if one allows both the metric and the tachyon to evolve with time, and cannot be seen if one studies the
brane decay process assuming  a flat space-time background. This is easy to see from the form of the effective metric. If
the scale factor $a$ is constant, the open string metric becomes (after a trivial rescaling of the spatial coordinates)
\begin{eqnarray}
\gb_{00}= -1+\dot{T}^2  \qquad \qquad \gb_{ij}=\delta_{ij} \, .
\end{eqnarray}
This metric is clearly flat because the coordinate transformation
\begin{eqnarray}
t^\prime(t) = \int^t ds\sqrt{1 - \dot{T}^2(s)}
\end{eqnarray}
will put the metric in the form $\gb_{\mu \nu}=\eta_{\mu \nu}$. The flatness of the effective metric can also be seen from
our expressions for the curvature tensor in Eq.~(\ref{curv}), which vanishes if $\dot{a}=\ddot{a}=0$. Thus, one would not
expect to see the large curvature corrections to the equations of motion in a flat spacetime background.  However, the
equation of state still becomes pressureless and the gauge couplings  become exponentially large   at late times even if
the background space-time is flat..

Recently, several authors have studied the decay of tachyonic branes in the presence of electric fields using Boundary
String Field Theory \cite{MS,RS} as well as effective field theory \cite{MS,RS,KKKK} techniques. These studies have found 
solutions in which the electric field is constant in time. It would be interesting to revisit this problem allowing for
a dynamical background spacetime. Our study shows that in these backgrounds the electric field can evolve in time,
though our approximations break down not long after the brane begins to decay. Corrections to the solution come from
nonlinear terms in the tachyonic Born-Infeld action, and Eq.~(\ref{eth}) shows that these nonlinearities should tame the
unbounded growth seen in our solution, Eq.~(\ref{et}). Another source of corrections to our results are higher derivative
corrections to the Born-Infeld action. Second and higher derivatives of the tachyon field are exponentially suppressed at
late times and so are under control. However, this is not true of higher derivatives of $E$ in the solution of
Eq.~(\ref{et}). However, the nonlinearities in the Born-Infeld action cut off the growth of $E$ so one may expect that 
terms involving derivatives of the field strength can be systematically neglected for solutions to the exact gauge field
and tachyon field equations. Finding explicit solutions to these equations  would help to address this issue but is beyond
the scope of the paper. A more serious problem  with  using a classical field theory based on the action in
Eq.~(\ref{ActionwithF}) is that as the D-brane decays as it becomes strongly coupled to massive string modes \cite{OS}.  A
recent study \cite{RS} shows that the presence of an electric field enhances the couplings of the decaying D-brane to
massive closed string states. These results show that the low energy effective action describing the decaying tachyon
should break down shortly after the brane begins to decay. Studying the behavior of decaying D-branes with electric fields
in a background space-time is clearly of interest.  However, to obtain a  reliable description one must go beyond the
effective field theory approach employed in this paper.

\section{Acknowledgments}
We thank N. Dalal, J. Gomis, B. Grinstein, K. Intriligator and J. Kumar for useful discussions. B.W. is supported in part by
DOE grant DE-FG03-97ER40546. T.M. is supported in part by DOE grants DE-FG02-96ER40945 and DE-AC05-84ER40150. T.M.
also acknowledges the hospitality of the University of California at San Diego Theory Group. 

\section{Appendix}

For reference, we note the following properties of the metric $\gb_{\mu \nu}$:
\vspace{.3cm}
The explicit form for the inverse effective metric $\gb^{\mu \nu}$ is
\begin{eqnarray}
\gb^{\mu \nu} = g^{\mu \nu} - \frac{\partial^{\mu} T \partial^{\nu}T}{1 + \partial_{\sigma} T \partial^{\sigma} T}.
\end{eqnarray}
In the following we will take $g_{\mu \nu}$ to be of FRW form with $k=0$ and consider time dependent, spatially
homogeneous tachyon fields. The Christoffel symbols for the metric $\gb_{\mu \nu}$ are
\begin{eqnarray}
\overline{\Gamma}^{i}_{0j} = \frac{\dot{a}}{a}\delta^{i}_{\,\, j}; \quad
\overline{\Gamma}^{0}_{ij} = \frac{1}{1-\dot{T}^2}\frac{\dot{a}}{a}\gb_{ij}; \quad
\overline{\Gamma}^{0}_{00} = \frac{-\dot{T}\ddot{T}}{1-\dot{T}^2}.
\end{eqnarray}
Using the convention $\left [ \nb_{\mu} , \nb_{\nu}\right ] V^{\alpha} = \rb^{\alpha}_{\,\,\, \beta \mu \nu}V^{\beta}$,
the nonzero components of the Riemann curvature tensor are
\begin{eqnarray}\label{curv}
\rb^{0}_{\,\,\, i0j} = \frac{1}{1- \dot{T}^2} \left [ \frac{\ddot{a}}{a} + \frac{\dot{a} \dot{T} \ddot{T}}{a (1-\dot{T}^2)} \right ]
\gb_{ij};
\quad
\rb^{i}_{\,\,\, jkl} = \frac{\dot{a}^2}{a^2 (1-\dot{T}^2)} \left ( \gb_{jl}\delta^{i}_{\,\,\, k} -
 \gb_{jk}\delta^{i}_{\,\,\, l} \right ).
\end{eqnarray}
The nonzero components of the Ricci tensor are thus
\begin{eqnarray}
\rb_{00} = -3 \left [ \frac{\ddot{a}}{a} + \left ( \frac{\dot{a}}{a} \right )
\frac{\dot{T}\ddot{T}}{1-\dot{T}^2} \right ]; \quad
\rb_{ij} = \frac{1}{1-\dot{T}^2} \left [ \frac{\ddot{a}}{a} + 2\left ( \frac{\dot{a}}{a} \right )^2 +
\left ( \frac{\dot{a}}{a} \right )\frac{\dot{T}\ddot{T}}{1-\dot{T}^2} \right ] \gb_{ij}
\end{eqnarray}
and the Ricci scalar is
\begin{eqnarray}
\rb = \frac{6}{1-\dot{T}^2} \left [ \frac{\ddot{a}}{a} + \left ( \frac{\dot{a}}{a} \right )^2
+ \left ( \frac{\dot{a}}{a} \right ) \frac{ \dot{T}\ddot{T}}{1-\dot{T}^2} \right ]. 
\end{eqnarray}

In deriving the energy-momentum tensor $T_{\mu \nu} = \frac{-2}{\sqrt{-g}}\frac{\delta {\cal L}}{\delta g^{\mu \nu}}$, the following 
identities are useful:

\begin{eqnarray} 
\frac{\delta \sqrt{-\gb}}{\delta g^{\mu \nu}} = -\frac{1}{2}\sqrt{-\gb}\gb^{\lambda \rho}g_{\lambda \mu}g_{\rho \nu} =
\frac{1}{2}\sqrt{-g}\sqrt{1+\partial_{\beta} T \partial^{\beta} T}
\left ( \frac{\partial_{\mu} T \partial_{\nu} T}{1+\partial_{\beta} T \partial^{\beta} T} - g_{\mu \nu} \right )
\end{eqnarray}
and
\begin{eqnarray}
\frac{\delta \gb^{\alpha \beta}}{\delta g^{\mu \nu}} = \gb^{\alpha \lambda} g_{\lambda \mu} \gb^{\beta \rho} g_{\rho \nu}.
\end{eqnarray}


\end{document}